\documentclass[%
 reprint, 
 amsmath,amssymb,
 aps, physrev,
floatfix,
]{revtex4-2}

\usepackage[english]{babel} 
\usepackage{xcolor}
\usepackage{url}
\usepackage[free-standing-units=true]{siunitx}
\usepackage{graphicx} 

\renewcommand{\eqref}[1]{Equation~\ref{#1}}
\newcommand{\figref}[1]{Fig.~\ref{#1}}
\newcommand{\figureref}[1]{Figure~\ref{#1}}
\newcommand{\figsref}[1]{Figs.~\ref{#1}}
\usepackage{dcolumn}
\usepackage{bm}

\begin{document}

\preprint{APS/}

\title{\textbf{On-Chip Frequency Noise Cancellation\\ in Nanomechanical Resonators using Cavity Optomechanics} 
}%

\author{Bhavesh Kharbanda}
\altaffiliation{Equal Contribution}
\affiliation{Laboratory of Solid State Physics, Dept. of Physics, ETH Zürich, Switzerland} 
\author{Amirali Arabmoheghi}
\altaffiliation{Equal Contribution}
\affiliation{Laboratory of Photonics and Quantum Measurements, EPFL, Switzerland} 
\author{Letizia Catalini}
\affiliation{Laboratory of Solid State Physics, Dept. of Physics, ETH Zürich, Switzerland} 
\affiliation{Center for Nanophotonics, AMOLF, The Netherlands}
\author{Mohammad Bereyhi}
\affiliation{Laboratory of Photonics and Quantum Measurements, EPFL, Switzerland} 
\author{Geena Benga}
\affiliation{Laboratory of Solid State Physics, Dept. of Physics, ETH Zürich, Switzerland} 
\author{Alessio Zicoschi}
\affiliation{Laboratory of Photonics and Quantum Measurements, EPFL, Switzerland} 
\author{Christian L. Degen}
\affiliation{Laboratory of Solid State Physics, Dept. of Physics, ETH Zürich, Switzerland}
\author{Tobias J. Kippenberg}
\affiliation{Laboratory of Photonics and Quantum Measurements, EPFL, Switzerland}
\author{Alexander Eichler}
\affiliation{Laboratory of Solid State Physics, Dept. of Physics, ETH Zürich, Switzerland} 
\author{Nils J. Engelsen}
\affiliation{Laboratory of Photonics and Quantum Measurements, EPFL, Switzerland} 
\email{nils.engelsen@epfl.ch}


\begin{abstract}

Understanding and minimizing the sources of frequency noise in nanomechanical resonators is crucial for many sensing applications. In this work, we report an ultracoherent perimeter-mode nanomechanical resonator co-integrated with an on-chip optical cavity. This device combines low thermomechanical force noise and low detector noise, allowing us to study its intrinsic frequency fluctuations in detail. We find that the fluctuations of two mechanical modes are strongly correlated; this holds not only for thermal drifts but also for intrinsic frequency flicker noise. Moreover, we demonstrate the generation of a signal at the frequency difference between the two modes directly on chip via nonlinear optomechanical transduction. This `difference signal' has vastly reduced intrinsic frequency fluctuations and can be used for frequency tracking with high precision, as we establish in a proof-of-principle experiment.
\end{abstract}
\maketitle

\emph{Introduction.—} Precise tracking of mechanical eigenfrequencies is central to most applications of micromechanical and nanomechanical systems, including clocks, filters, and accelerometers~\cite{kim2007frequency, Schmid2016FundamentalsResonators}, as well as sensors for scanning force microscopy~\cite{AFM1986, MFM1987,albrecht_frequency_1991, Rugar2004-MRFM}, mass spectroscopy~\cite{Naik2009mass_sensing,chaste2012nanomechanical}, pressure sensing~\cite{Reinhardt2023Calibration-lessRange}, and photothermal detection of nanoparticles~\cite{chien2018single}. In such applications, the measurement precision is limited by several sources of uncertainty~\cite{Cleland2002NoiseResonators,leeson2016oscillator,maizelis2011detecting, Rubiola2023TheVariances}: First, the thermomechanical motion of a resonator creates phase noise~\cite{Cleland2002NoiseResonators, Fong_2012_Frequency} in the oscillation with a power spectral density (PSD) equal to $S_\phi^\mathrm{th}(\Omega)$, where $\Omega$ is the sideband frequency, i.e., the offset from the resonance frequency $\omega_0$. When $\omega_0$ is being measured, this phase noise leads to a `flat' frequency noise PSD $S_\omega^\mathrm{th}(\Omega)\propto \Omega^0$. This noise sets a fundamental limit for the precision of frequency estimation at all timescales. Second, 
uncertainty in the frequency estimation is added by the readout (detection) noise. This noise leads to an \textit{apparent} frequency noise $S_\omega^\mathrm{det}(\Omega)\propto \Omega^2$, which typically dominates the estimation of $\omega_0$ at short times. Third, coupling of the resonator to two-level systems~\cite{Seoanez2008SurfaceElectrodes, Gruenke-Freudenstein2025TLSSurfaceResonators} induces intrinsic frequency flicker, also known as jitter~\cite{Fong_2012_Frequency, Sansa2016FrequencyNanoresonators}, whereas coupling to an environment with temperature changes causes frequency drifts and fluctuations~\cite{ZhangSt-Gelais-thermorefractive_mech2023}, owing to a finite thermal capacitance~\cite{Cleland2002NoiseResonators}. Together, both sources lead to a frequency noise PSD $S_\omega^\mathrm{int}(\Omega)\propto 1/\Omega^{\alpha}$, with $\alpha$ between $0.5$ and $2$. This \textit{real} frequency noise typically dominates at long times.

Various strategies have been used to optimize frequency stability and its estimation. The thermomechanical contribution $S_\omega^\mathrm{th}$ is minimized by designing devices with low dissipation, which was one of the central efforts in the nanomechanics community over the last decade~\cite{Tsaturyan2017-Patterened-membrane, Ghadimi2018-Corrugated-Strings, Fischer2019-UCB-RegalGroup, Bereyhi2022-Fractal, Bereyhi2022-Perimeter, Shin2022-Spider_Web_Ricard_Norte_HighQ}. Reducing the readout noise $S_\omega^\mathrm{det}$ requires greater detection sensitivity, which can be achieved with cavity optomechanics~\cite{AspelmeyerKippenberg2014, Gavartin2012A, Rossi2018Measurement, MohammadThesis2022, Guo_integrated_2022, gisler2024enhancing}. By contrast, overcoming intrinsic frequency fluctuations $S_\omega^\mathrm{int}$ is more challenging, as their microscopic origin is less well understood and harder to manipulate. In addition, this contribution is not amenable to time averaging as it impacts the measurement on long timescales. The influence of intrinsic frequency fluctuations can be reduced by subtracting the correlated fluctuations of various modes in postprocessing~\cite{Gavartin2013StabilizationLimit}. However, such analysis is generally susceptible to uncorrelated detector noise and cannot always be applied to real-time measurements. Alternatively, one can drive a nonlinear resonator at an optical working point~\cite{yurke1995theory,Kenig2012PassiveScheme,villanueva2013surpassing,Brown2026High-StabilityRegime}, which requires careful calibration and fine-tuning of the system. Until now, no method has been shown that offers on-chip, calibration-free cancellation of frequency fluctuations. 

In this work, we address this challenge with an on-chip optomechanical platform. The device is an ultralow-dissipation perimeter-mode nanomechanical resonator~\cite{Bereyhi2022-Perimeter} co-integrated with an on-chip optical cavity~\cite{MohammadThesis2022}. With low thermomechanical and readout noise, we can study the intrinsic frequency fluctuations over several orders of magnitude in $\Omega$. Importantly, we find that the fluctuations measured on two different mechanical modes are strongly correlated. Taking advantage of the system's nonlinear optomechanical transduction, we generate a signal at the frequency difference of the modes directly on the chip. This signal has vastly reduced frequency fluctuations and can be used as a resource for precise frequency tracking. We demonstrate this functionality in a proof-of-principle frequency measurement experiment.

\begin{figure*}[!tbp]
 \includegraphics[]{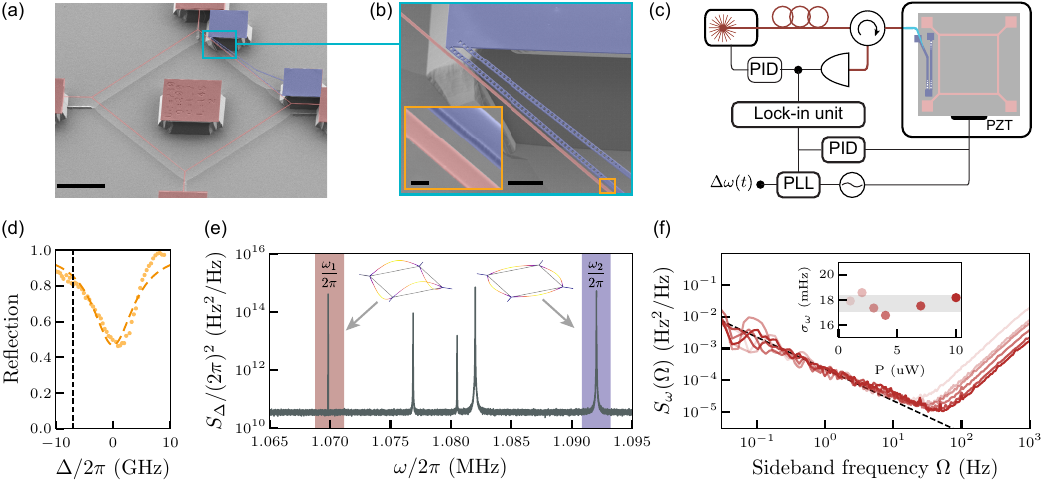}
 \caption{Measurement setup and calibration. (a)~False-color scanning electron microscope micrograph of the Si$_3$N$_4$ optomechanical system. Red: Polygon resonator. Blue: Fabry P\'erot cavity (scalebar $\SI{50}{\micro\meter}$). (b)Detail showing the photonic crystal mirrors (scalebar $\SI{5}{\micro\meter}$). Inset: Polygon resonator suspended near the cavity (scalebar $\SI{200}{\nano\meter}$). Nanobeam-waveguide gap: \SI{200}{\nano\meter}. (c)~Schematic of the optical detection scheme. $\Delta\omega$ = measured frequency shift, see Fig.~\ref{fig:3}. (d)~Normalized reflection trace of the optical mode used in the experiment. Yellow circles: measured reflection as a function of laser wavelength. Yellow dashed line: fit of the optical reflection. Vertical dashed line: laser operating point at $\Delta = $\SI{-7}{\giga\hertz}, used in experiments. (e)~Displacement PSD of the calibrated optical frequency noise. The peaks correspond to the thermomechanical noise stemming from the perimeter modes of the polygon resonator. Modes marked with red and blue are used in the experiment. (f)~Mechanical frequency noise PSD of the high-$Q$ mode at different optical powers. Inset: Allan variance corresponding to the $A/\Omega$ component of the PSDs as a function of the input optical power. The gray shading shows the standard deviation of all values. PZT = piezoelectric actuator, PID = feedback loop, PLL = phase-locked loop}
 \label{fig:1}
\end{figure*}

\emph{Experimental platform.—} The optomechanical system is formed by co-integration of a polygon nanomechanical resonator~\cite{Bereyhi2022-Perimeter} with a photonic crystal optical micro-cavity. The polygon resonator is suspended in the near field of the cavity, and its motion couples dispersively to the optical modes through the evanescent field.~\figsref {fig:1}(a) and (b) show scanning electron microscope micrographs of a device. The devices are fabricated from \SI{250}{\nano\meter}-thick stoichiometric silicon nitride (Si$_3$N$_4$) with \SI{1.1}{\giga\pascal} deposition stress. The polygon resonator is partially thinned down to \SI{100}{\nano\meter}. The devices are then patterned by electron beam lithography and dry etching, and are finally released with the use of a KOH wet process. The details of the design and fabrication of these devices are discussed in Ref.~\cite{MohammadThesis2022}.

The polygon resonator used in our experiment is composed of a square with \SI{245}{\micro\meter}-long and \SI{500}{\nano\meter}-wide sides, which is suspended with \SI{49}{\micro\meter}-long tethers. The high stress and high aspect ratio of these devices allow us to exploit dissipation dilution~\cite{Fedorov2019GeneralizedResonators}, enhancing the quality factor $Q$ of the flexural modes by a factor of up to $10^6$~\cite{Engelsen2024Ultrahigh,beccari_visani_kippenberg_NatPhys2022, Bereyhi2022-Fractal, Bereyhi2022-Perimeter}. The resonator hosts a family of modes confined to the perimeter of the device, which we refer to as `perimeter modes'. These modes feature especially high quality factors due to the soft-clamping effect~\cite{Tsaturyan2017-Patterened-membrane, Bereyhi2022-Perimeter}.
The optical cavity is realized by etching holes in the two ends of a \SI{250}{\nano\meter}-wide single-mode Si$_3$N$_4$ waveguide, forming a Fabry-P\'erot cavity~\cite{MohammadThesis2022} operating at telecom wavelengths. The cavity is driven and read out via an auxiliary reflector waveguide (single-sided cavity). The polarization-sensitive waveguide is optically addressed using a tip-etched optical fiber, see supplementary information (SI)~\ref{SI:cavity} for details.

\figureref{fig:1}(c) illustrates our optical detection scheme. Continuous-wave laser light from a Toptica CTL 1550 laser is to drive the photonic crystal cavity at a fixed cavity detuning. We probe a cavity resonance at \SI{1552.6}{\nano\meter} with total linewidth $\kappa/2\pi$ = \SI{7.66}{\giga\hertz}; see~\figref{fig:1}(d). The cavity has a free spectral range of \SI{4.25}{nm} (see Supplemental Material~\ref{SI:cavity} for the data and calibration~\cite{albert-schliesser-thesis2009}). Its dc component is used to lock the laser frequency \SI{5.97}{\giga\hertz} red-detuned from the cavity resonance, while the ac component encodes the nanobeam’s motion. To minimize gas damping, the chip is placed in a vacuum chamber at a pressure of about $\SI{1e-7}{\milli\bar}$. A piezoelectric actuator, clamped to the chip, is used to excite the mechanical modes of the polygon resonator. 

We detune the laser by $\Delta = \SI{5.97}{\giga\hertz}$ to the red side of the resonance and observe mechanical motion through direct detection. Using a digital lock-in amplifier (Zurich Instruments MFLI), we measure the PSD of the mechanical displacement driven by thermomechanical force noise, see~\figref{fig:1}(e). The peaks correspond to the thermal motion of the mechanical modes of the polygon resonator, as confirmed by FEM simulations (see insets). This PSD is calibrated in units of frequency fluctuations of the optical resonance, see Supplemental Material~\ref{SI:freq_noise} for details. The mechanical modes shown in this frequency range correspond to the high-$Q$ fundamental out-of-plane (OOP) perimeter mode and a family of in-plane (IP) modes, also confined to the perimeter of the device.
The two modes used in our experiments are the OOP mode with frequency $\omega_1$ = \SI{1.070}{\mega\hertz} and the IP mode with frequency of $\omega_2$ = \SI{1.092}{\mega\hertz}. Using the ringdown technique, we measure quality factors of $Q_1 = 4\times 10^7$ and $Q_2 = 3\times 10^5$ for the OOP and IP modes, respectively. For these modes, we also calibrate the vacuum optomechanical coupling rates to be $g_{0,1}/ 2\pi = \SI{5.2}{\kilo\hertz}$ and $g_{0,2}/ 2\pi = \SI{15.6}{\kilo\hertz}$ (see Supplemental Material~\ref{fig:ringdowns} for ringdown measurements and details of the calibration).

We use a digital phase-locked loop (PLL) to measure the modes' frequencies as a function of time. The frequency noise PSD of the OOP mode shown in~\figref{fig:1}(f) is measured for different input optical powers at the same detuning from the cavity resonance, shown as a dashed gray line in~\figref{fig:1}(d). Because of the high $Q$, the thermomechanical contribution is too small to be measured, and the spectra can be fit well with a model composed of only the $1/\Omega$-component shown as a dashed gray line in \figref{fig:1}(f) in addition to detection noise contributions. We clearly see power-dependent detection noise at large sideband frequencies, while $1/\Omega$-type fluctuations at low sideband frequencies are independent of power. Independence of these fluctuations from mechanical drive power, pressure~\cite{yang_surface_2011} and temperature fluctuations is also shown in Supplemental Material~\ref{SI:f-noise-independence}. We use the $A/\Omega$ component of the spectra to calculate the corresponding Allan deviation for long integration times as $\sigma_\omega = \sqrt{2A\ln{(2)}}$, where $A$ is a constant (see Supplemental Material~\ref{SI:f-noise-calib} for details). These values are shown in the inset in~\figref{fig:1}(f), showing the power independence. These measurements confirm that the frequency noise for long integration times (low sideband frequencies) is dominated by $1/\Omega$-type fluctuations that are not related to optical absorption. 


\begin{figure*}[!tbp]
 \includegraphics[]{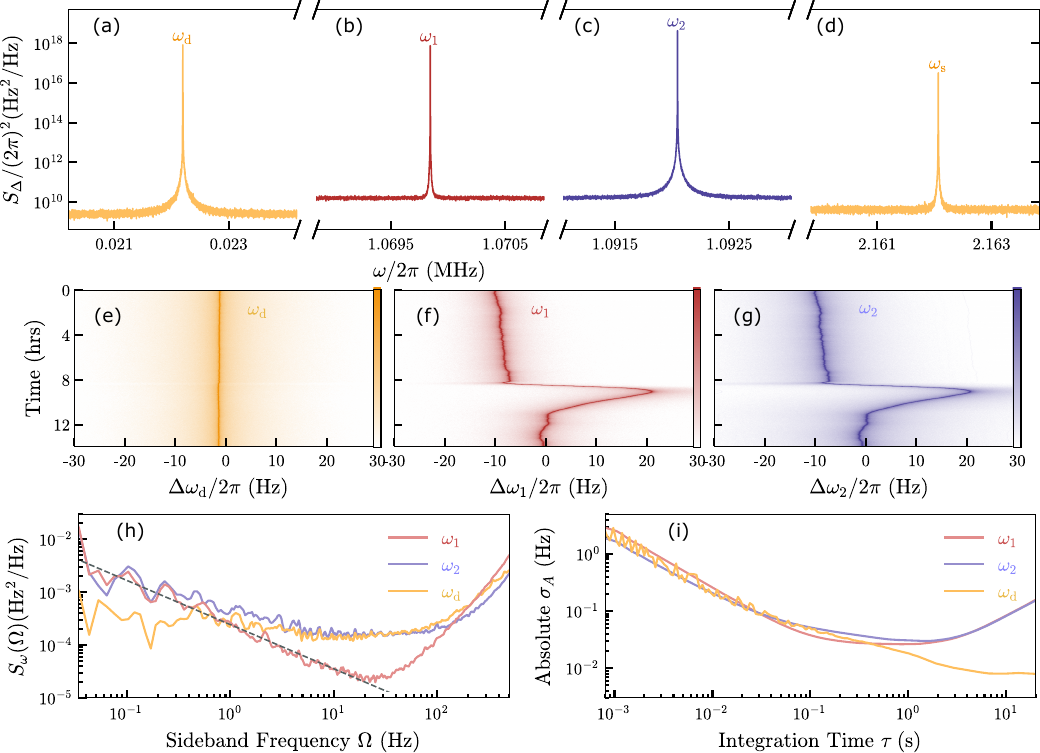}
 \caption{Properties of signal at $\omega_\mathrm{d} = \omega_2 - \omega_1$. Displacement PSD of demodulated signal around (a)~$\omega_\mathrm{d} = \omega_2 - \omega_1$ (b)~$\omega_1$, (c)~$\omega_2$, and (d)~$\omega_s = \omega_1 + \omega_2$. The mechanical modes at $\omega_1$ (red) and $\omega_2$ (blue) are driven with separate PLLs. (e)-(g)~Spectograms of the difference signal and the displacements of the OOP and the IP modes. The colour bar corresponds to the normalized PSD. (h)~Double-sided frequency-PSD for $\omega_1$, $\omega_2$, and $\omega_\mathrm{d}$ spanning \SI{150}{\second} with a data rate of \SI{14.3}{\kilo\hertz}. We use the PLLs for estimating $\omega_1$ and $\omega_2$, while $\omega_\mathrm{d}$ is calculated using phase-to-frequency conversion~\cite{Besic2023ResonanceResonators}. The dashed-gray line is the fit for the $1/\Omega$-component of the frequency noise PSD. (i)~Absolute Allan deviation $\sigma$ in \SI{}{\hertz} calculated using the same traces as in (h).}
 \label{fig:2}
\end{figure*}

\emph{Nonlinear frequency noise cancellation.—} Interestingly, when $\omega_1$ and $\omega_2$ are driven simultaneously, we observe optical signatures at the sum and difference frequencies ($\omega_{d} = \omega_2 - \omega_1$, $\omega_{s} = \omega_2 + \omega_1$), see~\figref{fig:2}(a)-(d). These signatures are due to the nonlinear optical transduction of the mechanical signal by the cavity (see Supplemental Material~\ref{SI:nonlinear_transduction} for details) that leads to mixing of the two signals~\cite{Fedorov2020ThermalMeasurements}. More precisely, the cavity detuning is a time-varying parameter due to the mechanical motion of the resonator. This detuning modulates the intensity of the out-coupled cavity response, where the first-order signal is at the mechanical resonance frequencies, and the second-order effects are at their linear combinations.
The displacement PSDs in Figs.~\ref{fig:2}(a)-(d) provide direct access to various frequency signal components. However, the $1/\Omega$-type frequency fluctuations are not easily visible in such a PSD on short timescales.
To visualize the frequency fluctuations, we monitor how the voltage PSDs change over time, see Figs.~\ref{fig:2}(e)-(g). Indeed, we find that both $\omega_1$ and $\omega_2$ change by more than \SI{30}{\hertz}---hundreds of linewidths for the OOP mode---over several hours. The fluctuations of the two modes are strongly correlated but exhibit no clear pattern as a function of time, as expected for $1/\Omega$-type fluctuations. We also observe strong dependence on temperature changes (see Supplemental Material~\ref{SI:thermal-drifts},\cite{land_sub-ppm_2024}). By contrast, the signal at $\omega_\mathrm{d}$ is significantly more stable, with a maximum change of about \SI{0.4}{\hertz}, roughly two orders of magnitude smaller than that of $\omega_{1,2}$.

We illustrate the improvement in the corresponding frequency noise PSD (with uniform distribution of points along the x-axis~\cite{Trobs_Heinzel_LPSD_paper, BodeChristoph_LPSD_Python}) in Fig.~\ref{fig:2}(h). The two modes at $\omega_1$ and $\omega_2$ follow the same $1/\Omega$ slope, shown as a dashed gray line. The difference mode, however, has much lower contributions at sideband frequencies below \SI{1}{\hertz}, marked by a decrease by an order of magnitude in the $1/\Omega$-type branch. We observe that the noise floor is limited by the thermomechanical noise contribution of the mode with lower $Q$, i.e. $\omega_2$, to around \SI{1.4e-4}{\hertz^2/\hertz}. In the corresponding Allan deviation $\sigma$ in Fig.~\ref{fig:2}(i), $\sigma_{\omega_1}$ is at its minimum value $\SI[separate-uncertainty = true]{26.15 (0.03)}{\milli\hertz}$ around $\SI{0.9}{\second}$ and starts increasing again for longer times. The difference signal, however, has a $\sigma_{\omega_\mathrm{d}}$ up to a factor of 20 lower than for the OOP and IP modes at long integration times, reaching $\SI[separate-uncertainty = true]{7.54(0.01)}{\milli\hertz}$ at $\tau = \SI{25}{\second}$. 
This effect is directly tied to the origin of this `difference signal' as a mixing product of modes $\omega_1$ and $\omega_2$. As the frequency fluctuations of the two real modes are correlated, their frequency difference remains approximately constant. As we show in Appendix D in the Supplementary Material~\ref{SI:freq_noise}, the correlated fluctuations stem both from environmental temperature changes and intrinsic flicker noise, and are both reduced by our optomechanical mixing technique. Both sources can be identified separately and can be shown to follow different power laws for frequencies beyond \SI{0.01}{\hertz}.



\begin{figure*}[!t]
  \begin{minipage}[t]{0.75\textwidth}
  \vspace{0pt}
    \includegraphics[width = \textwidth]{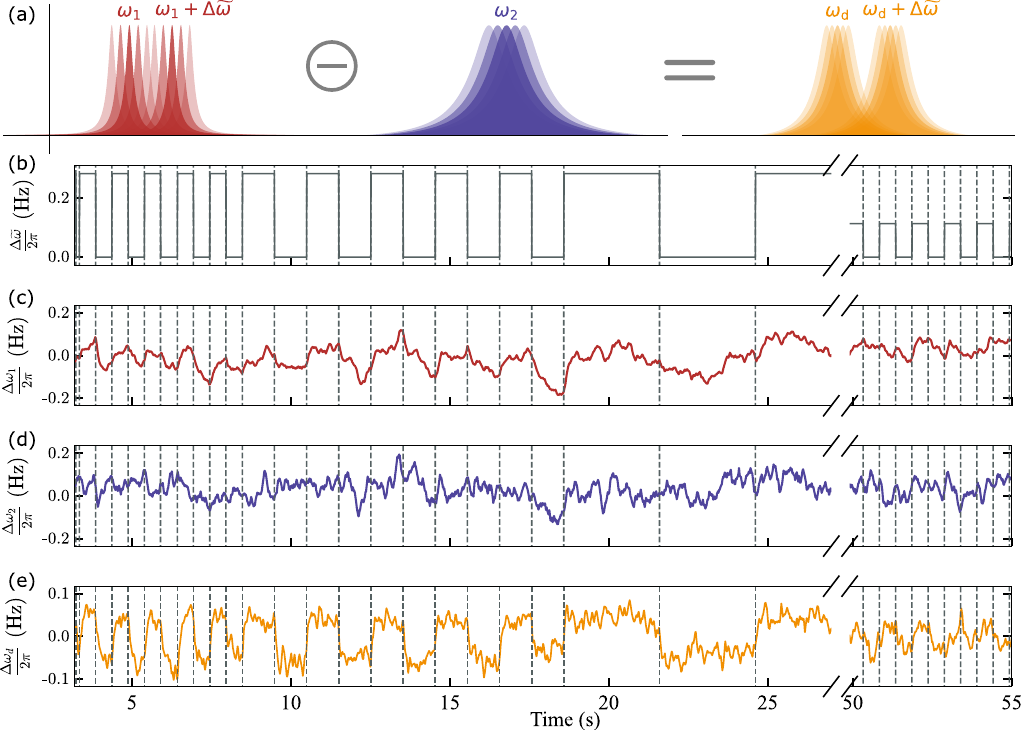}
  \end{minipage}\hfill
  \begin{minipage}[t]{0.2\textwidth}
  \vspace{0pt}
  \parfillskip=0pt
  \leftskip=0pt
  \rightskip=0pt
\noindent Figure 3. Measuring frequency jumps with high precision at $\omega_\mathrm{d}$. (a) Effect of frequency jumps $\Delta\omega$ acting on $\omega_1$, but not on $\omega_2$. (b)~Expected frequency jumps $\Delta\widetilde{\omega}$ when a dispersive feedback on mode $\omega_1$ is toggled on and off. The two modes at $\omega_1$ and $\omega_2$ are driven by separate PLLs. Frequency shifts (c)~$\Delta\omega_1$, (d)~$\Delta\omega_2$, and (e)~$\Delta\omega_\mathrm{d}$ are simultaneously measured as a function of time.
  \end{minipage}
  \label{fig:3}
\end{figure*}

\emph{Feedback frequency shift experiment.—} In Fig.~\ref{fig:3}, we demonstrate that the stable signal at $\omega_\mathrm{d}$ is potentially useful for frequency measurements. As in Fig.~\ref{fig:2}, we simultaneously drive and measure $\omega_1$ and $\omega_2$ with a PLL, while demodulating the signal at $\omega_\mathrm{d}$ without applied drive. Additionally, we use an active feedback loop tuned to induce a pure frequency shift $\Delta\widetilde{\omega}$ on the OOP mode at $\omega_1$ without affecting the mechanical linewidth (by applying a feedback in phase with the oscillator position, resulting in softening of the spring constant~\cite{poggio2007feedback,kleckner2006sub, Stomp2024ResonanceInstruments}). Toggling this feedback signal on and off results in frequency jumps of $\omega_1$ without affecting $\omega_2$. We refer to the frequency shift signal using the normalized gain parameter $U\in[0,1]$ where $U = 1$ corresponds to the maximum $\Delta\widetilde{\omega}$ shown in the graph. This controllable frequency modulation is used to simulate the frequency jumps expected, for example, in single-particle mass sensing experiments~\cite{Naik2009mass_sensing,chaste2012nanomechanical}.

In the time trace of $\omega_1$ in Fig.~\ref{fig:3}, it is difficult to distinguish the response to $U$ by eye, as it is partially masked by frequency fluctuations. As expected, $\omega_2$ does not display any systematic response to $U$ at all. For a quantitative analysis, we refer to each section between the vertical gray dashed lines as a block. We calculate the signal-to-noise ratio (SNR) as our metric, where we use the mean frequency jump size between two blocks as the signal and the standard deviation within each block to quantify the noise. 
We compare the SNR of the measured $\omega_\mathrm{d}$ against the measured $\omega_1$ and the value $\omega^\mathrm{cal}_\mathrm{d} = \omega_2-\omega_1$ calculated in postprocessing~\cite{Gavartin2013StabilizationLimit}. For unprocessed data, we obtain an SNR of $4.1 \pm 0.3$ at $\omega_\mathrm{d}$ in contrast with an SNR of $2.4 \pm 1.7$ at $\omega_1$ and an SNR of $1.3 \pm 0.1$ for $\omega^\mathrm{cal}_\mathrm{d}$, noting an increase by a factor of 3.2 at the same frequency. When we perform a simple moving-average filter over 200 data points in time (corresponding to a sync filter time of \SI{24}{\milli\second}), the SNR at $\omega_\mathrm{d}$ is $6.2 \pm 1$, while we obtain an SNR of $2.5 \pm 1.2$ at $\omega_1$ and $4.5 \pm 0.6$ for $\omega^\mathrm{cal}_\mathrm{d}$. The SNR of the difference signal is consistently better than that of the post-processed $\omega^\mathrm{cal}_\mathrm{d}$, as well as that of $\omega_1$ itself. In the SI, we present an alternative analysis based on cross-correlations, arriving at a similar result.

We assign the increased SNR of the signal at $\omega_\mathrm{d}$ relative to that measured at $\omega_1$ to the cancellation of intrinsic frequency fluctuations at low $\Omega$. As we see in Fig.~\ref{fig:2}(h), this effect is significant up to roughly $\Omega=\SI{1}{\hertz}$, explaining the reduced frequency fluctuations and the improved SNR in Fig.~\ref{fig:3}. However, for $\Omega>\SI{1}{\hertz}$, we can see in Fig.~\ref{fig:2}(h) that the signal at $\omega_\mathrm{d}$ has a greater $S_\omega$ than $\omega_1$ due to the increased thermomechanical force noise that it inherits from $\omega_2$. This noise contribution can be mitigated by averaging the data over time. This is why the SNR at $\omega_\mathrm{d}$ increases from 4.1 to 6.2 when the moving-average filter is applied, while the one at $\omega_1$ remains constant within the measurement uncertainty. In principle, the low-$\Omega$ fluctuations can also be reduced in postprocessing by calculating the signal $\omega^\mathrm{cal}_\mathrm{d} = \omega_2-\omega_1$~\cite{Gavartin2013StabilizationLimit}. However, the SNR of this signal always remains below that of $\omega_\mathrm{d}$ because it samples the detection noise twice, once when $\omega_1$ is measured and once for $\omega_2$. Assuming that the detection noise imprinted on the two PLL measurements is uncorrelated, the resulting high-$\Omega$ noise present for $\omega^\mathrm{cal}_\mathrm{d}$ should be roughly $\sqrt{2}$ times larger than that for $\omega_\mathrm{d}$. This is in agreement with the observation that averaging increases the SNR at $\omega^\mathrm{cal}_\mathrm{d}$ significantly. Nevertheless, it never quite catches up with the on-chip noise cancellation we demonstrate with $\omega_\mathrm{d}$.

\emph{Conclusions and outlook.—} In this work, we perform frequency noise characterization of high-$Q$ devices and observe correlated intrinsic frequency fluctuations of different mechanical modes. This correlation for low-mass~\cite{ZhangSt-Gelais-thermorefractive_mech2023}, high-$Q$ resonators is expected to predominantly be correlated $1/\Omega$-noise~\cite{Fong_2012_Frequency}, resulting from the shared defects interacting with the various mechanical modes. Using nonlinear optomechanical transduction, we show how the correlation can be used for on-chip frequency noise reduction. With this method, we achieve a 20-fold decrease in absolute Allan deviation at times larger than $\SI{1}{\second}$. Furthermore, from comparison of the SNRs, the method performs up to a factor $3.2$ better than off-chip frequency noise reduction by postprocessing, which is attributed to the method not being limited by PLL errors in the frequency measurements of individual modes, as the detection noise is sampled only once, not independently for both modes. Finally, we verify the usability of this noise suppression method in a proof-of-principle experiment that mimics the conditions for mass sensing. We find similar behavior across multiple devices and by engaging different modes. Therefore, we conclude that our method can be a useful resource for precise frequency sensing experiments, such as mass spectroscopy~\cite{Naik2009mass_sensing,chaste2012nanomechanical} or photothermal particle detection~\cite{chien2018single}. High-precision frequency tracking, combined with the excellent force sensitivity and low detection noise of these perimeter-mode resonators, is also crucial for nuclear spin detection~\cite{Visani2026Near-resonantResonators,kovsata2020spin}, and ultimately for nanoscale magnetic resonance imaging~\cite{budakian2024roadmap}.

In our demonstration, the limiting factor for frequency noise up to sidebands of \SI{100}{\hertz} is the thermomechanical force noise of the IP mode at $\omega_1$, whose quality factor is significantly lower than that of the OOP mode at $\omega_2$. Alternative resonator designs with two high-$Q$ modes~\cite{Catalini2020, Bereyhi2022-Perimeter} can potentially overcome this problem and achieve even better performance. For instance, with two modes that both possess $Q=5 \times 10^7$, the total measured frequency variance would be reduced from \SI{134}{\milli\hertz} to \SI{78}{\milli\hertz}, with a frequency noise PSD below \SI{3e-5}{\hertz^2/\hertz} over sideband frequencies up to \SI{100}{\hertz}. Such a device will exhibit consistently low frequency noise PSD ($S_\omega$) over a large range of sideband frequencies, serving as a precise time reference at both short and long timescales.

\emph{Acknowledgments.} This research was possible thanks to the Swiss National Science Foundation (SNSF) through Grants No. 200021\_200412 and No. 200021E\_209235, and the Novo Nordisk Foundation through Grant No. NNF22OC0077964. N.J.E. acknowledges funding from the European Research Council (ERC) under Grant No. 101117144. We would like to thank Albert Schliesser (Niels Bohr Institute, University of Copenhagen), Silvan Schmid (TU Vienna), Hajrudin Bešić (TU Vienna), and Olivier Faist (Zurich Instruments) for insightful discussions and nudges; Diego Visani, Thomas Gisler, Nils Prumbaum, Urs Grob and Luis Mestre are acknowledged for in-laboratory assistance. The technical support from Walter Bachmann of the engineering office and from the mechanics workshop of the Department of Physics at ETH Zürich, is highly appreciated.

\emph{Data Availability.—} The data that support the findings of this article are openly available.~\cite{DataPRAppl2026}.

\bibliography{arXiV-full-text.bbl}
\newpage
\begin{appendix}

\onecolumngrid
\begin{center}
  {\large\textbf{Supplemental Material for:}\\
  {On-Chip Frequency Noise Cancellation in Nanomechanical Resonators\\
  using Cavity Optomechanics}}
\end{center}
\twocolumngrid

\vspace{20pt}
\section{Cavity Characterisation}\label{SI:cavity}
\begin{figure*}[!tbp]
 \includegraphics[]{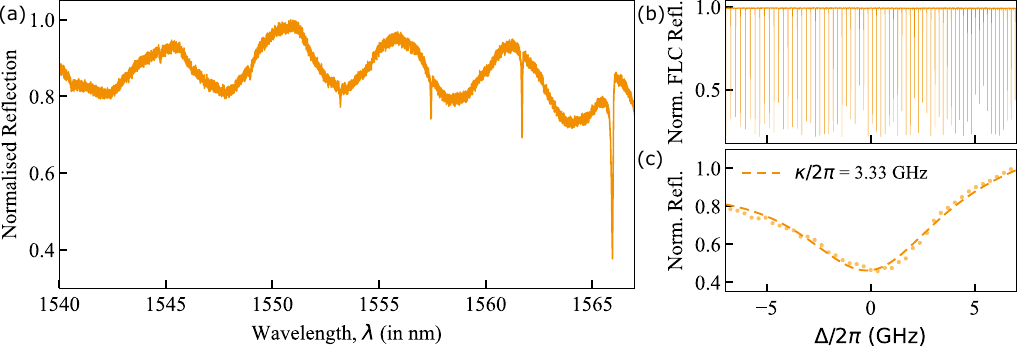}
 \caption{Wide cavity scan and narrow scan with calibration. (a)~Wide-range reflection spectrum of the cavity showing multiple resonances with a FSR of \SI{4.25}{nm}.  (b)~Reflected signal of a fibre loop cavity (FLC) used for calibrating the x-axis  (c)~Reflected power from the rightmost cavity mode in (a) as a function of laser detuning $\Delta/2\pi$. $\Delta = 0$ indicates the cavity resonance. The fit in orange dashed yields a linewidth of $\kappa/2\pi = 3.33$ GHz.}
 \label{fig:cavity_wide}
\end{figure*}

We use a sharply tapered fibre, with a full tapering angle of about 6~$\deg$, for input coupling to the integrated optical cavity. We optimize the fibre position on top of the auxiliary waveguide and the polarization of the incoupled light before the optical circulator to maximize the reflected optical power. We then perform a wide-range wavelength scan of the light reflected back from the integrated optical cavity. We observe multiple cavity modes with a free-spectral range (FSR) of \SI{4.25}{nm}. We select the mode at \SI{1566}{\nano\meter} as the working mode. After using the fine scan function to record a high-resolution scan of the mode of interest, Fig. \ref{fig:cavity_wide} (c)  we lock to the red side of the cavity using the built-in cavity lock feature of the Toptica laser.  The calibration of the x-axis is performed using a fibre-loop cavity (FLC)~\cite{albert-schliesser-thesis2009}.  

\section{Optomechanical Characterisation}
\begin{figure*}[!tbp]
 \includegraphics[]{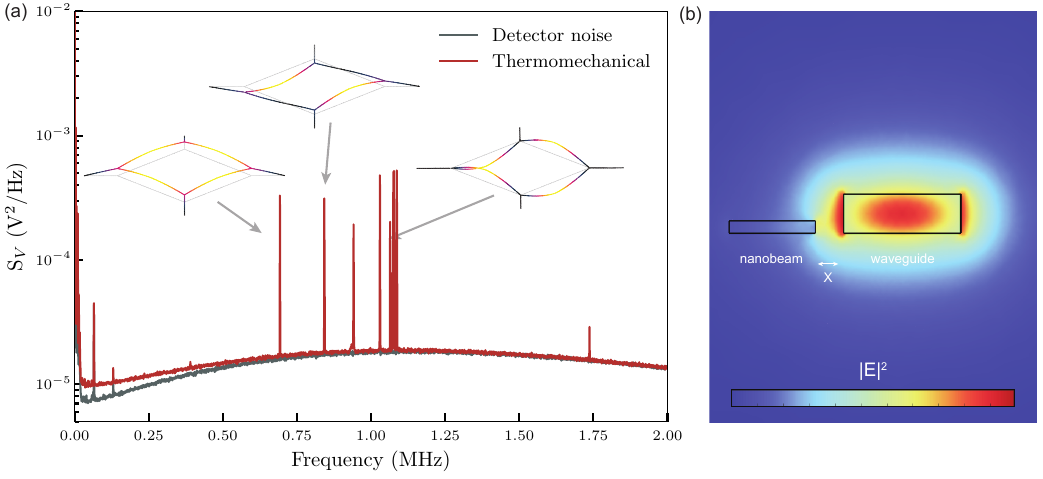}
 \caption{Optomechanical coupling. (a) A broadband spectrum obtained in direct detection from the device. The red line corresponds to the thermomechanical spectrum acquired when the laser is red detuned from the cavity. The peaks correspond to the thermal motion of different modes of the polygon resonator, a few of which are marked alongside their mode shapes. (b) FEM simulation of the electric field intensity }
 \label{fig:Broadband_spectrum}
\end{figure*}

We characterise the mechanical properties of the perimeter resonator by first measuring a thermal spectrum as shown in~\figref{fig:Broadband_spectrum}(a), at the red-detuned sideband of the optical cavity. The first and second order modes of the resonator as well as the fundamental perimeter mode alongside the FEM simulation  of their mode shapes are shown in the figure.~\figref{fig:Broadband_spectrum}(b)~shows an FEM simulation of the cross-section the optical mode in the waveguide in the presence of the nanobeam. From this simulation we can infer the derivative of the guided refractive index with respect to the gap (i.e. $dn_\mathrm{eff}/dX)$. One can calculate the optomechanical pulling factor using the relation $G = \omega_c \frac{1}{n_\mathrm{eff}}\frac{dn_\mathrm{eff}}{dX}$. For our device, we estimate this quantity to be about \SI{0.5}{\giga\hertz/\nano\meter} for the OOP perimeter mode, in agreement with our estimation of the optomechanical coupling rate.

Following this, we drive the modes at $\omega_1$ and $\omega_2$ using a PLL feeding back on a piezo actuator. We then perform a ringdown for each mode to characterise the $Q$ of each mode as shown in main text Fig. 1(e).

\begin{figure*}[!tbp]
 \includegraphics[]{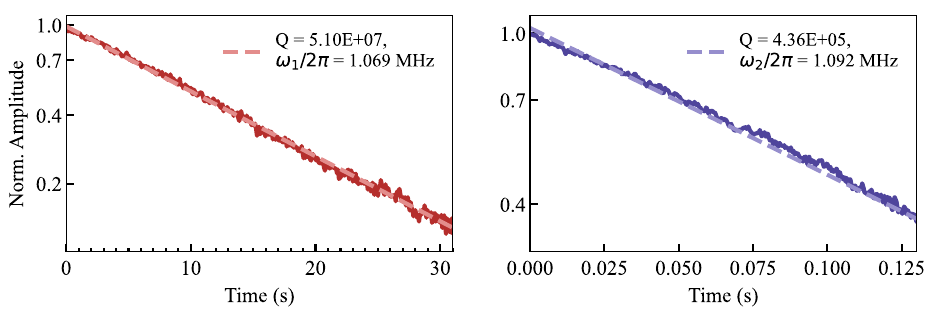}
 \caption{Ringdown measurements of the device's perimeter modes $\omega_1$ in red and $\omega_2$ in blue with their respective fits (dashed). }
 \label{fig:ringdowns}
\end{figure*}
\section{Optical frequency noise calibration}\label{SI:f-noise-calib}
In this section the calibration procedure for optical cavity frequency and vacuum optomechanical coupling rates is explained. Within the linear transduction regime, when the laser is detuned from the cavity resonance, in direct detection the photocurrent is given by
$$ I(t) = D^{(0)} +  D^{(1)}\xi(t),$$
where
\begin{align}
    D^{(0)} &= 1 - \frac{4\eta(1-\eta)}{1 + \bar\delta^2},\\
    D^{(1)} &= 4\eta(1-\eta)\frac{2\bar\delta}{(1 + \bar\delta^2)^2},
   \end{align}
and $\xi(t)$ is the cavity frequency noise normalized to $\kappa/2$, containing the mechanical displacement. See later in SI for the derivations. The technique used here is based on using the to use the DC component of the photocurrent to calibrate the AC component. The DC and AC components of the detected voltage is given by
\begin{align}
    V_\mathrm{DC} &= Z_\mathrm{DC}D^{(0)},\\
    V_\mathrm{AC}(t) &= Z_\mathrm{AC}D^{(1)}\xi(t),
   \end{align}
where $Z_\mathrm{DC}$ and $Z_\mathrm{AC}$ are values of the trans-impedance gain at DC and AC frequencies. For the power spectral density (PSD) of the AC signal, defined as $S_{VV}(\omega) = \int e^{i\omega\tau}\langle V_\mathrm{AC}(t)V_\mathrm{AC}(t+\tau)\rangle d\tau$, it can be written as
\begin{equation}\label{eq:S_VV_calibration}
    S_{VV}(\omega) = \left[ \frac{Z_\mathrm{AC}}{Z_\mathrm{DC}} \right]^2\left[\frac{V_\mathrm{DC}D^{(1)}}{D^{(0)}}\right]^2 S_{\xi\xi}(\omega).
\end{equation}
Without knowing the exact value of the trans-impedance gains and only knowing the ratio $\frac{Z_\mathrm{AC}}{Z_\mathrm{DC}}$, one can fully calibrate the spectrum. Assuming the detector has a flat response and $\frac{Z_\mathrm{AC}}{Z_\mathrm{DC}}=1$ the conversion factor for voltage to detuning noise, $S_{\Delta\Delta} = C^2 S_{VV}$, is given by
$$ C = \frac{\kappa}{2} \left[ \frac{D^{(0)}}{V_\mathrm{DC}D^{(1)}} \right]\quad (\mathrm{Hz}/\mathrm{V}) $$
This conversion is used to calibrate the data presented in the main-text Fig. 1(e). To calibrate the vacuum optomechanical coupling rate $g_0$ one has to integrate both side of Eq. \ref{eq:S_VV_calibration} around a mechanical sideband. Integration of $S_{\xi\xi}$ around a mechanical mode with frequency $\omega_m$ and OM coupling $g_0$ obtains $8g_0^2n_\mathrm{th}/k^2$ where $n_\mathrm{th} = k_BT/\hbar\omega$ is the thermal occupation of the mode. Using this, $g_0$ is given by
\begin{equation}
    g_0 = \frac{CV_\mathrm{rms}}{\sqrt{2n_\mathrm{th}}},
\end{equation}
where $V_\mathrm{rms}$ is the RMS value of the mechanical sideband in the voltage spectrum.
\section{Frequency noise characterization}\label{SI:freq_noise}
When measuring mechanical frequency fluctuations, it is known that the measurement is contaminated by three main sources of noise: thermomechanical, detection, and intrinsic frequency noise, 

\begin{equation}
    \delta\omega(t) = \delta\omega_\mathrm{th}(t) + \delta\omega_\mathrm{det}(t) + \delta\omega_\mathrm{int}(t).
\end{equation}
These sources are statistically independent from each other, such that the power spectral density (PSD) of the total frequency noise is given by: 
\begin{equation}\label{eqn:f-PSD}
    S^\text{tot}_\omega(\Omega) =  S_\omega^\mathrm{th}(\Omega) + S_\omega^\mathrm{det}(\Omega) + S_\omega^\mathrm{int}(\Omega),
\end{equation}
 where the PSD is defined as the Fourier transform of the correlation function
 \begin{equation}
     S_\omega^a(\Omega) = \int_\mathrm{\infty}^\infty\langle\delta\omega_a(0)\delta\omega_a(\tau)\rangle e^{i\omega\tau}d\tau,
 \end{equation}
 for $a = \mathrm{th}, \mathrm{det}, \mathrm{int}$. The detection noise originates from the noise added to the signal at the photodetection stage whereas the thermomechanical and intrinsic noise are properties of the mechanical resonator.
 
 For the difference frequency, $\omega_\text{d} = \omega_2 - \omega_1$, we can construct the noise model based on this model to give,
 \begin{multline}
     \delta\omega_\text{d}(t) = \delta\omega_\mathrm{th,2}(t) - \delta\omega_\mathrm{th,1}(t) + \delta\omega_\mathrm{det,d}(t)\\\delta\omega_\mathrm{int,2}(t) - \delta\omega_\mathrm{int,1}(t).
 \end{multline}
 Note that while the thermomechanical and intrinsic components stem from the corresponding components in $\omega_1$ and $\omega_2$, the detection component, however, appears only once. This is because this difference frequency tone is generated within the optomechanical cavity, and prior to detection. Therefore, the contribution does not get summed in.
 
 The corresponding frequency noise PSD is given by
 \begin{multline}
     S_{\omega_\text{d}} = S_\omega^\mathrm{th,1}(\Omega) + S_\omega^\mathrm{th,2}(\Omega)+
     S_\omega^\mathrm{det,d}(\Omega)\\
     S_\omega^\mathrm{int,1}(\Omega) + S_\omega^\mathrm{int,2}(\Omega) - 2S_\omega^\mathrm{int,12}(\Omega),
 \end{multline}
 where
  \begin{equation}
     S_\omega^\mathrm{int,12}(\Omega) = \int_\mathrm{\infty}^\infty\langle\delta\omega_\mathrm{int,1}(0)\delta\omega_\mathrm{int,2}(\tau)\rangle e^{i\omega\tau}d\tau,
 \end{equation}
 corresponds to the correlation between the intrinsic noise components. 
 
 We fit these models to our experimental data to extract the frequency noise properties. It is known that the thermomechanical term has no frequency dependence and that the detection term has a quadratic frequency scaling. For the data shown in the main-text Fig. 1(f), we use 
 the model $S_\omega(\Omega) = B\Omega^2 + A/\Omega$ for fitting and extracting the $1/\Omega$ noise component. We have not included a constant term since, as mentioned in the main text, the thermomechanical component for the high-$Q$ OOP mode is too low to be resolved. 
 
 All frequency PSD calculations are done using the linearised LPSD python package \url{https://pypi.org/project/lpsd/}. In addition, the SciPy Cross-PSD package \url{https://docs.scipy.org/doc/scipy/reference/generated/scipy.signal.csd.html} was used to calculate frequency noise correlations. 

\begin{figure}
 \includegraphics[]{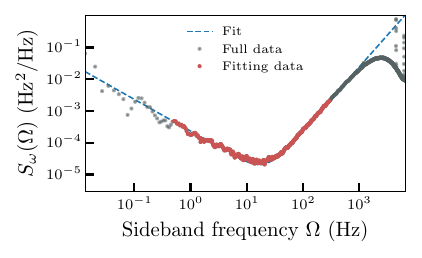}
 \caption{Example of the frequency noise model fitting for extraction of the $1/\Omega$ component coefficient for the lowest value of the input power.}
 \label{fig:FNoise_fitting}
\end{figure}

 In~\figref{fig:FNoise_fitting} we show an example of the full frequency noisePSD, the segment used for fitting, and the result of the fit for the lowest input optical power. The roll-off at frequencies above \SI{1}{\kilo\hertz} is a result of the finite bandwidth of the lock-in amplifier's demodulator. We do not include this segment in the model. The initial points of the data are not included either due to the lack of sufficient number of points and large variations. $1/\Omega$-type frequency noise with a PSD $S_\omega(\Omega) = A/\Omega$ corresponds to a constant Allan variance with standard deviation $\sigma^2_\omega(\tau) = 2A\ln{(2)
 }$. This relation is used to produce the inset of the main-text Fig. 1(f).
 
Furthermore we observe and characterise the strongly correlated nature of $1/\Omega$ noise in the two modes $\omega_1$ and $\omega_2$. We do so by evaluating the frequency noise PSD (Eqn: \ref{eqn:f-PSD}) of both modes ($S_{\omega_1}$ and $S_{\omega_2}$ respectively) and then calculate the cross-PSD, given by
 \begin{equation}
     S_{\omega_a\omega_b}(\Omega) = \int_\mathrm{\infty}^\infty\langle\delta\omega_a(0)\delta\omega_b(\tau)\rangle e^{i\omega\tau}d\tau,
 \end{equation}.
 
\begin{figure} 
 \includegraphics[]{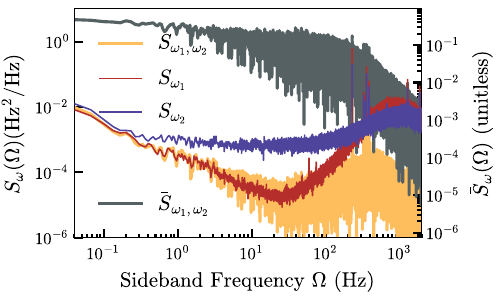}
 \caption{frequency noise PSD of the mechanical modes $\omega_1$ (red) and $\omega_2$ (blue) as shown in main-text Fig: 2(h). Cross-PSD, $S_{\omega_1\omega_2}$ in yellow and the Correlation f-PSD normalised to the geometric mean ($\bar{S}_{\omega_1\omega_2} = S_{\omega_1\omega_2}/\sqrt{S_{\omega_1}S_{\omega_2}}$) of the individual PSDs in grey}
 \label{fig:FNoise_corr}
\end{figure}

 It is worth noting in \figref{fig:FNoise_corr} that $S_{\omega_1\omega_2}(\Omega)$ in yellow follows the same trend as that of $S_{\omega_1}(\Omega)$ in red. This signifies that this $1/\Omega$ contribution from both modes $\omega_1$ and $\omega_2$ is maximally correlated. As an alternate visualisation, we also plot $\bar{S}_{\omega_1\omega_2}(\Omega) = S_{\omega_1\omega_2}(\Omega)/\sqrt{S_{\omega_1}(\Omega)S_{\omega_2}(\Omega)}$, which is normalised to the point-by-point individual PSDs. It shows that the signals are correlated to nearly the same extent throughout the entire bandwidth, $\Omega$ of up to \SI{100}{\hertz}.

\subsection{Impact of the measurement conditions on the $1/\Omega$ noise}
\begin{figure*}[!tbp]
 \includegraphics[]{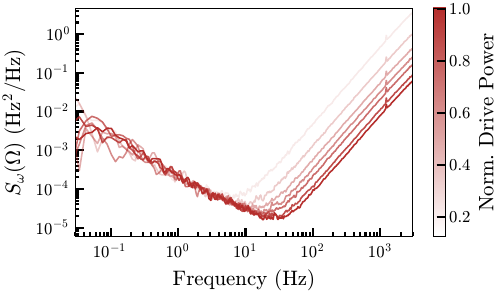}
 \includegraphics[]{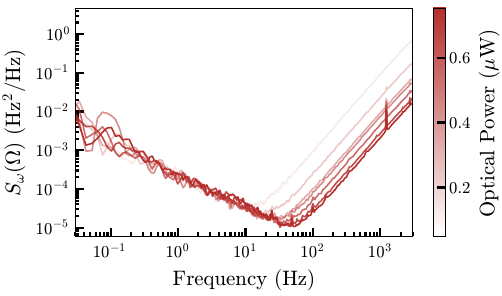}
 
 \includegraphics[]{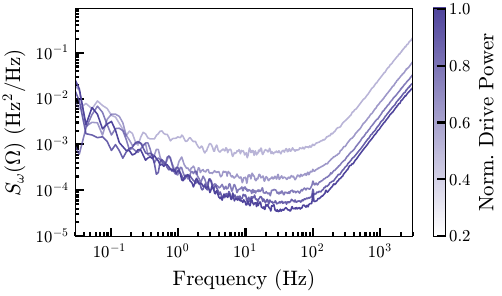}
 \includegraphics[]{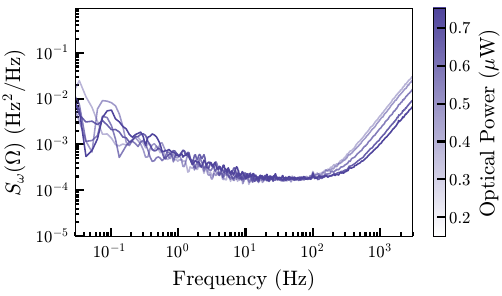}

 \caption{Measured frequency power spectral density of modes $\omega_1$ and $\omega_2$ for different driving powers (left) and optical power at fixed drive power(right). Similar to the effect of varying the optical power circulating in the cavity, we see that the $1/\Omega$ noise branch remains consistent across all measurements, while the high-frequency response changes with the driving power.}
 \label{fig:Sup_optical_drive_pressure_Fnoise}
\end{figure*}

In the mechanical frequency fluctuations measurements, the two components that mainly affect the noise floor are measurement optical power and mechanical driving amplitude. It is expected that increasing the driving amplitude, reduces the thermomechanical and the detector contributions and increasing the optical power reduces the detector contribution as it increases the measurement SNR~\cite{Besic2023ResonanceResonators}. We verify these properties experimentally by the measuring the frequency noise PSDs for both modes for different optical powers and driving amplitudes. These measurements are shown in \figref{fig:Sup_optical_drive_pressure_Fnoise}. As expected, we observe that the mechanical driving power affects both thermomechanical and detector contributions (note that for the high-Q mode the thermomechanical noise is not resolved as it is much weaker than the other contributions). We also observe that increasing the optical power consistently improves the detection noise. The other important observation is that neither optical power nor the driving power affects the $1/\Omega$ noise contribution for either of the modes. The independence of the $1/\Omega$ noise on the measurement conditions, particularly the optical power, is the first evidence for the $1/\Omega$ noise to be intrinsic to the resonator.

\section{Distinguishing the intrinsic $1/\Omega$ noise from environmental effects}\label{SI:f-noise-independence}
\subsection{Thermal drifts}\label{SI:thermal-drifts}
\begin{figure*}[!tbp]
 \includegraphics[]{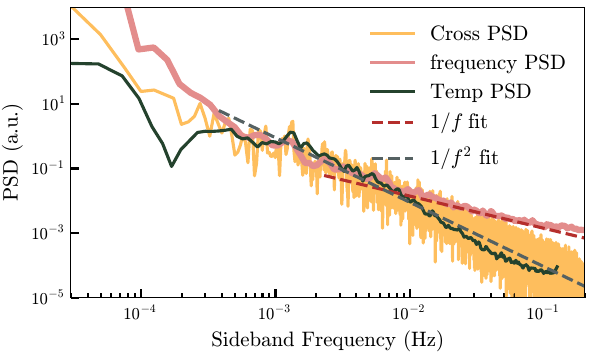}
 \caption{Frequency noise PSD of the mechanical mode $\omega_1$ in comparison with the Temperature PSD, both normalised to arbitrary units. We see that the temperature noise branch has a much faster decay than the $1/\Omega$ branch at shorter time scales, implying that the processes are only correlated for timescales slower than $\SI{1e-2}{\hertz}$.}
 \label{fig:Sup_FvsT}
\end{figure*}
The observed frequency fluctuations at low sideband frequencies $\Omega$ have two possible causes: temperature fluctuations of the device, either due to environmental temperature drift or laser intensity fluctuations, and the intrinsic frequency flicker noise. In the following, we show that the observed frequency fluctuations can be ascribed to flicker noise alone.

Temperature variations in the environment are transduced to the polygon resonator through multiple thermal pathways, each characterized by a thermal frequency response of the generic form
\begin{equation}
    R_\mathrm{th}(\Omega) = (1 + \Omega^2\tau_\mathrm{th}^2)^{-1}
\end{equation}
where $\tau_{\mathrm{th}}$ is the characteristic thermal response time of a given body. For the chip and the assembly on which it is mounted, $\tau_{\mathrm{th}}$ is on the order of seconds~\cite{land_sub-ppm_2024}, while for the polygon resonator, the thermal time constant is about \SI{13.5}{\milli\second}, assuming thermal conduction occurs primarily through the tethers. Consequently, temperature fluctuations originating outside the chip are low-pass filtered by the thermal response of the chip, suppressing fluctuations faster than $\sim$1 Hz. However, in~\figref{fig:Sup_optical_drive_pressure_Fnoise} we clearly observe a $1/\Omega$-power law well above 1 Hz. These frequency fluctuations cannot stem from the environment. 

To visualize the difference between regimes of thermal fluctuations and flicker noise, we performed simultaneous measurements of the mechanical resonance frequency and the vacuum chamber temperature over a 7-hour period. The normalized PSDs of both quantities are shown in~\figref{fig:Sup_FvsT}. For frequencies below $10^{-2}$ Hz, both PSDs follow a $1/\Omega$ scaling and are strongly correlated. 

In order to explain the $1/\Omega$-type frequency fluctuations above 1 Hz, we consider heating sources that act locally on the device. The most likely source is intensity fluctuations of the laser light arriving on the chip. To test whether this source is responsible for the observed fluctuations, we performed frequency fluctuation measurements with different laser powers. As shown in~\figref{fig:Sup_optical_drive_pressure_Fnoise}, the different laser powers strongly affect the readout noise beyond 100 Hz, but leave the $1/\Omega$-type noise unchanged. For this reason, we rule out laser intensity fluctuations as the source of our frequency fluctuations between 1 and 100 Hz.

As both environmental and local temperature fluctuations are excluded as possible sources for the observed frequency fluctuations between 1 and 100 Hz, we consider intrinsic frequency flicker noise as the dominant source. Flicker noise is believed to stem from coupling of the resonator to two-level systems in the material and is expected to contribute to frequency fluctuations~\cite{Fong_2012_Frequency}. Clearly, we have evidence that temperature fluctuations are dominant in this regime. By contrast, above $10^{-2}$ Hz the PSD power laws decouple, with the frequency fluctuation PSD following a $1/\Omega$ law. This provides clear evidence that the mechanical frequency fluctuations at these higher frequencies arise from a non-thermal mechanism.

\subsection{Gas pressure}
\begin{figure}
 \includegraphics[]{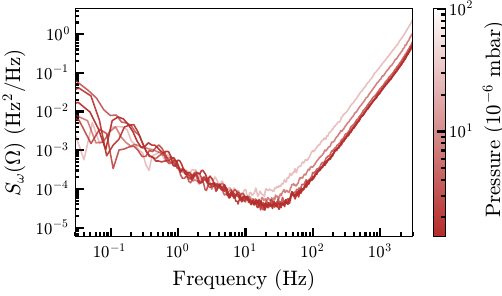}
 \caption{Measured mechanical frequency noise PSD for $\omega_1$ for varying vacuum chamber pressures in the range $\SI{1e-6}{}$ to $\SI{1e-4}{\milli\bar}$}
 \label{fig:Sup_pressure_Fnoise}
\end{figure}

Adsorption of gas particles on the surface of the resonator could also lead to mechanical frequency .~\cite{yang_surface_2011}. To study the impact of this phenomenon we measure the frequency noise of the high-Q mode as a function of pressure. As shown in \figref{fig:Sup_pressure_Fnoise}, no significant change is observed for pressures below $\SI{1e-4}{\milli\bar}$, implying it does not influence the intrinsic $1/\Omega$ noise. The change in the high frequency, detector-noise-limited regime is due to the reduction of the mechanical $Q$ of this mode due to gas damping and the resulting reduced oscillation amplitude.

\section{Optical Non-linear Transduction for difference Frequency generation}\label{SI:nonlinear_transduction}
\begin{figure}
 \includegraphics[]{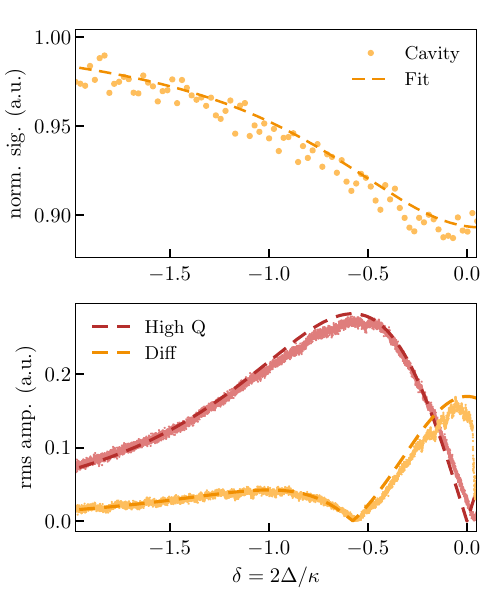}
 \caption{(Above) Left-half (red-detuned side) of the optical cavity resonance shown in~\figref{fig:cavity_wide}. (Below) Transduced signal for mode $\omega_1$ (High Q mode) and the signal at $\omega_\text{d}$ (Diff signal). the latter shows a decay in signal at the magic detuning, $\bar\delta = 1/\sqrt{3}$~\cite{Fedorov2020ThermalMeasurements} }
 \label{fig:Sup_intermod}
\end{figure}

\begin{figure*}[!tbp]
 \includegraphics[]{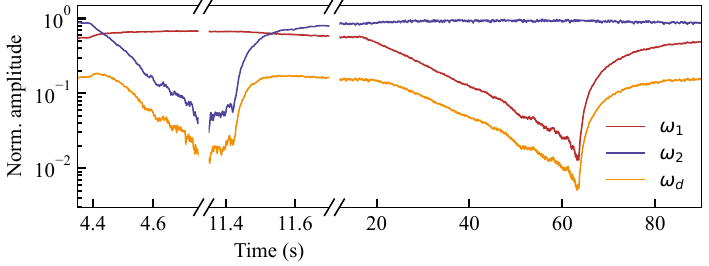}
 \caption{
 Simultaneous ringdown measurements. Both mechanical modes at frequencies $\omega_1$ and $\omega_2$ are driven simultaneously, and the corresponding response at the frequency difference $\omega_\mathrm{d}$ is monitored. In the first sequence, the drive at $\omega_2$ is turned off while $\omega_1$ remains driven. A correlated decay and subsequent revival of both the $\omega_2$ mode and the signal at $\omega_\mathrm{d}$ is observed, indicating coupling between these modes. In the second sequence, the drive at $\omega_1$ is turned off, and a similar correlated decay is observed between $\omega_1$ and $\omega_\mathrm{d}$. 
 }
 \label{fig:Sup_cor_ring}
\end{figure*}

The presence of a difference (and sum) frequency signal in the reflected laser light comes from the non-linear transduction of the mechanical signal by the photonic cavity. To derive the said dependence, we begin by studying the photon-mechanics interaction.
The equation of motion of the intra-cavity field $a(t)$ in an optical cavity of linewidth $\kappa$ at a detuning $\Delta$ is given by
\begin{equation}
  \dot a = \left(i\Delta(t) - \frac{\kappa}{2}\right)a + \sqrt{\kappa_\mathrm{ext}}a_\mathrm{in}
\end{equation}
We assume that the detuning is modulated by two mechanical modes driven to amplitudes $x_1(t)$ and $x_2(t)$ respectively
\begin{equation}
  \Delta(t) = \bar\Delta + G_1x_1(t) + G_2x_2(t)
\end{equation}
We further assume that the mechanical modes are strongly driven (compared to the thermomechanical drive), and hence
\begin{equation}\label{eqn:x_motion}
  x_i = X_i\cos\Omega_it + X_{th}(t) \approx X_i\cos\Omega_it
\end{equation}
and that we are well inside the sideband-unresolved regime, $\kappa \gg \Omega_1,\Omega_2$. With this assumption, the optical field adiabatically follows the detuning fluctuations. The intra-cavity field can therefore be expanded as
\begin{equation}
  a(t) = \frac{2\sqrt{\kappa_\mathrm{ext}}a_\mathrm{in}}{\kappa}\frac{1}{1 - i \frac{2\Delta(t)}{\kappa}}.
\end{equation}
We can define the normalized detuning, the cavity coupling efficiency and the normalised amplitude as
\begin{align}
  \delta(t) &:= \frac{2\Delta(t)}{\kappa}\\
  \bar\delta &:= \frac{2\bar\Delta}{\kappa}\\
  \eta &:= \kappa_\mathrm{ext}/\kappa\\
 \xi(t) &:= \frac{2G_1x_1}{\kappa} + \frac{2G_2x_2}{\kappa}  
\end{align}
Using the input-output relation
\begin{equation}
  a_\mathrm{out} = a_\mathrm{in} - \sqrt{\kappa_\mathrm{ext}}a,
\end{equation}
we can find the field reflected from the cavity
\begin{equation}
  a_\mathrm{out} = a_\mathrm{in}\left[ 1 - \frac{2\eta}{1-i\delta(t)} \right]
\end{equation}
In a direct detection scheme, the photocurrent is expressed as
\begin{equation}
  I(t) = \frac{|a_\mathrm{out}|^2}{|a_\mathrm{in}|^2} = \left|1 - \frac{2\eta}{1-i\delta(t)}\right|^2 = 1 - \frac{4\eta(1-\eta)}{1 + \delta^2}.
\end{equation}
and then expand $I(t)$ to second order in $\xi$
\begin{equation}
\begin{split}
   I(t) \approx 1 &- \frac{4\eta(1-\eta)}{1 + \bar\delta^2} + \frac{4\eta(1-\eta)2\bar\delta}{(1+\bar\delta^2)^2}\xi(t) \\ &+ \frac{4\eta(1-\eta)(1-3\bar\delta^2)}{(1+\bar\delta^2)^3}\xi(t)^2
\end{split}
\end{equation}
We define the transduction coefficients to different orders of $\xi$ 
\begin{align}
  D^{(0)} &= 1 - \frac{4\eta(1-\eta)}{1 + \bar\delta^2}\\
  D^{(1)} &= \frac{4\eta(1-\eta)2\bar\delta}{(1+\bar\delta^2)^2}\\
  D^{(2)} &= \frac{4\eta(1-\eta)(1-3\bar\delta^2)}{(1+\bar\delta^2)^3}
\end{align}
The photocurrent is given by 
\begin{equation}\label{eqn:NL_Intensity}
  \begin{split}
    I(t) = &D^{(0)} + \frac{D^{(2)}}{2}\left( \delta_1^2 + \delta_2^2\right)+\\
        &\frac{D^{(2)}}{2}\delta_1\delta_2\cos (\Omega_2 - \Omega_1)t+\\
        &D^{(1)}\delta_1\cos\Omega_1t+ D^{(1)}\delta_2\cos\Omega_2t+\\
        &\frac{D^{(2)}}{2}\left\{\delta_1^2\cos 2\Omega_1t + \delta_2^2\cos 2\Omega_2t 
        + \delta_1\delta_2\cos (\Omega_2+\Omega_1)t\right\}
  \end{split}
\end{equation}
with $\delta_i = 2G_iX_i/\kappa$. In particular, the difference frequency component is given by
\begin{equation}
  I_\mathrm{diff}(t) = \frac{D^{(2)}}{2}\delta_1\delta_2\cos (\Omega_2 - \Omega_1)t
\end{equation}
Since we are neglecting the thermomechanical noise (\eqref{eqn:x_motion}), all the signal terms appear as delta functions in the spectrum and the power of the difference frequency sideband is given by
\begin{equation}
    \begin{split}
  P_\mathrm{diff} &= \frac{{D^{(2)}}^2\delta_1^2\delta_2^2}{8} \\ &= \frac{32\eta^2(1-\eta)^2G_1^2G_2^2X_1^2X_2^2}{\kappa^2}\left[\frac{1-3\bar\delta^2}{(1+\bar\delta^2)^3}\right]^2, 
\end{split}
\end{equation}
  s shown in~\figref{fig:Sup_intermod}, one scans through the red-detuned sideband of the cavity and monitors a driven mode signal and the transduced signal difference. The dashed lines in~\figref{fig:Sup_intermod}(b) are a single free parameter fit, namely the signal amplitude. Evidently, the nonlinear transduction (the difference signal) vanishes at $\bar\delta = 1/\sqrt{3}$, termed the magic detuning introduced in Ref:\cite{Fedorov2020ThermalMeasurements}. Including the thermomechanical noise would exactly replicate the Ref:~\cite{Fedorov2020ThermalMeasurements}.
We also demonstrate the optical transduction nature of the signal with another experiment. The transduction difference signal is substantial when the mode frequencies  $\omega_1$ and $\omega_2$ are both driven. In~\figref{fig:Sup_cor_ring} we plot a symbolic signal 'Trigger' which is unity only when both modes are simultaneously driven, and zero otherwise. While monitoring both mode amplitudes and the difference signal amplitude, the parameter of interest is the decay time of each signal. When the drive to $\omega_1$ is switched off, the decay rate of $\omega_1$ and $\omega_\mathrm{d}$ follow the same trend. However, if the same is done while driving $\omega_1$ and switching the drive for $\omega_2$ off, the decay rate of $\omega_\mathrm{d}$ is defined by that of $\omega_2$, hence demonstrating the pure transduction nature of the signal which exhibits no intrinsic decay rate of its own.

\section{Detector Bandwidth}
In this section, we characterise the transfer function of our photodetection apparatus. We use a test characterisation setup with an EOM (Exail MPZ-LN-01 Intensity Modulator, operational range up to \SI{3}{\giga\hertz}) and a polarisation controller to modulate the laser light directly that is incident on the photodiode - transimpedance amplifier detection electronics. The transimpedance amplifier is expected to have an 8 dB smaller signal at \SI{2.16}{\mega\hertz} compared to that at \SI{22}{\kilo\hertz}, as per the provided specifications sheet. In order to experimentally validate this, we sweep a modulation tone across a range of \SI{10}{\kilo\hertz} to \SI{5}{\mega\hertz} and record the demodulated signal. We also send a modulation tone at frequencies of interest - $ \omega_\text{d} \sim $ \SI{22}{\kilo\hertz}, $\omega_\text{1} \sim $ \SI{1.06}{\mega\hertz} and  $\omega_\text{s} \sim $ \SI{2.16}{\mega\hertz} in successive experiments to plot the voltage Power Spectral Density. We, therefore, experimentally observe over 13 dB of signal drop compared to \SI{}{\kilo\hertz} frequency range where the detector response is rather flat. This largely explains our experimental observation in main-text Fig 2(a, d) where the the signal  at $\omega_\text{d}$ and $\omega_\text{s}$ has unequal signal strengths.
\begin{figure*}[!tbp]
 \includegraphics[]{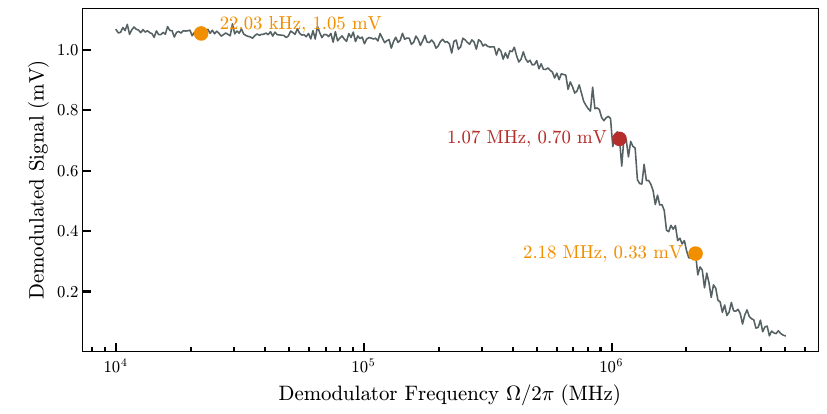}
 \includegraphics[]{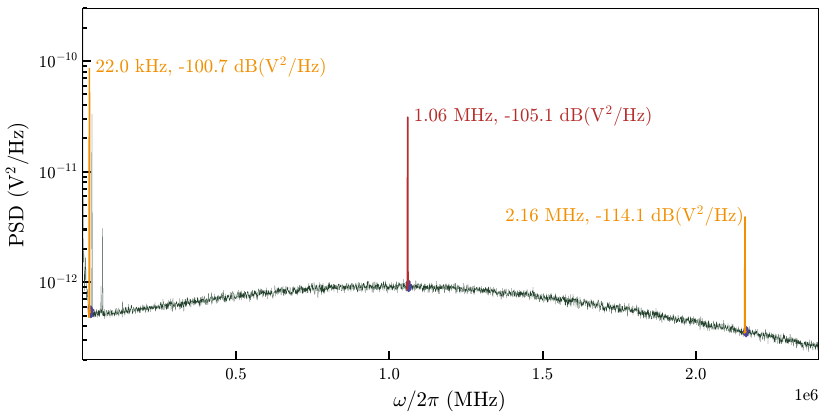}
 \caption{Characterization of the responsivity of the detection branch (photodiode and transimpedance amplifier) using an intensity modulator to (a) control the demodulated signal while sweeping the modulation frequency.(b) record the voltage PSD of the laser signal (in grey) and modulate sucessively  close to three frequencies of interest, - $\omega_\text{d} $, $\omega_\text{1}$ and  $\omega_\text{s}$.}
 
 \label{fig:Intensity_modulator}
\end{figure*}

\section{Data Acquisition and Processing}
We employ two (Zurich Instruments) MFLI units for all mentioned measurements. The first lock-in is used for driving the mechanics of the two modes $\omega_1$ and $\omega_2$ using a demodulator and a PLL for each with target BW \SI{2}{\kilo\hertz} for each of the two modes, running at a data rate of \SI{2.14e6}{Sa/s}. Two other demodulators are used for generating the feedback signal that softening and stiffens the spring constant of the mode $\omega_1$ used in the main-text Fig. 3. The proportional gain for the left and right sections is $20$ and $8$.  The physics and implementation are almost identical to that of feedback cooling~\cite{poggio2007feedback,kleckner2006sub}, the only difference being that the feedback force $F_{fb}$ is not in phase with the velocity $dx/dt$ of the mechanical mode but with the position, $x$, of the mechanical mode, hence $\pi/2$ phase-shifted relative to that for feedback cooling. It therefore introduces a frequency shift $\Omega_{fb}$.
\begin{eqnarray}
    m_{\mathrm{eff}} \frac{d x^2(t)}{d t^2}+m_{\mathrm{eff}} \Gamma_m \frac{d x(t)}{d t}+m_{\mathrm{eff}} \Omega_m^2 x(t)=F_{\mathrm{dr}}(t) + F_{\mathrm{fb}}(t) \nonumber\\
     m_{\mathrm{eff}} \frac{d x^2(t)}{d t^2}+m_{\mathrm{eff}} \Gamma_m \frac{d x(t)}{d t}+m_{\mathrm{eff}} \left[ \Omega_m^2 \pm \Omega_{fb}^2\right] x(t)=F_{\mathrm{dr}}(t) \nonumber
\end{eqnarray}

\section{2D drifts}

\begin{figure*}[!tbp] 
 \includegraphics[]{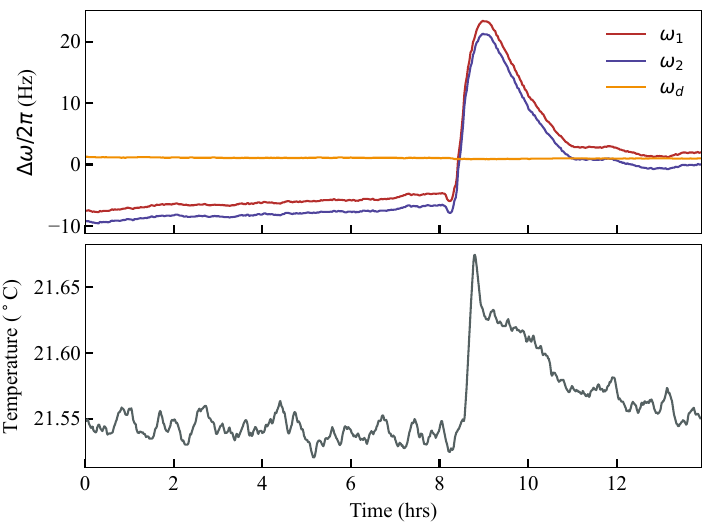}
 \caption{
Time evolution of the center frequency of the $\omega_1$, $\omega_2$ and $\omega_\mathrm{d}$ and the room temperature change over a period of 14 hours. The data shows a strong correlation in the temporal shifts of modes $\omega_1$ and $\omega_2$, while the signal at the frequency difference $\omega_\mathrm{d}$ remains stable and unaffected throughout the measurement period.
}
 \label{fig:Sup_2D_peaks}
 \end{figure*}
Measurement of these modes is done at room temperature. As seen in main text Fig. 2(e)-(g), the frequency changes suddenly around the \SI{8}{hr} mark. This is correlated with a change in temperature of the laboratory as shown in ~\figref{fig:Sup_2D_peaks} due to start of movement in the lab after a night of calm. 


\section{Sensing demonstration: Non-linear intrinsic cancellation vs post-processing subtraction}\label{sec:SI_mass_sensing}
We analyse the data shown in the main-text section IV. Sensing Experiment using SNR as our figure of merit. Here, we present an alternative quantitative analysis. We refer to the correlations between any two signals $s_1(t)$ and $s_2(t)$ -- we calculate their normalized correlation coefficient $\rho(s_1, s_2) = \langle s_1,s_2 \rangle/\sqrt{\langle s_1\rangle \langle s_2 \rangle}$, where $\langle...\rangle$ indicates an average over time, implying that 1 and 0 indicate fully correlated and uncorrelated signals, respectively. We refer to $U(t)$ as the feedback gain that causes an expected frequency shift $\Delta\widetilde{\omega}$. We use the non-averaged data for the below for immediately next discussion where the averaged data used for plotting the main-text Fig. 3 shows very similar numbers.

Calculating the normalised correlation signals, We find $\rho(U, \omega_1) = 0.53$ and $\rho(U, \omega_2) = 0.03$, confirming the visual impression. The correlated intrinsic frequency fluctuations of the two modes lead to a finite correlation $\rho(\omega_1, \omega_2) = 0.44$ between the modes, independently of $U$.

\begin{figure*}[!tbp]
  \includegraphics[]{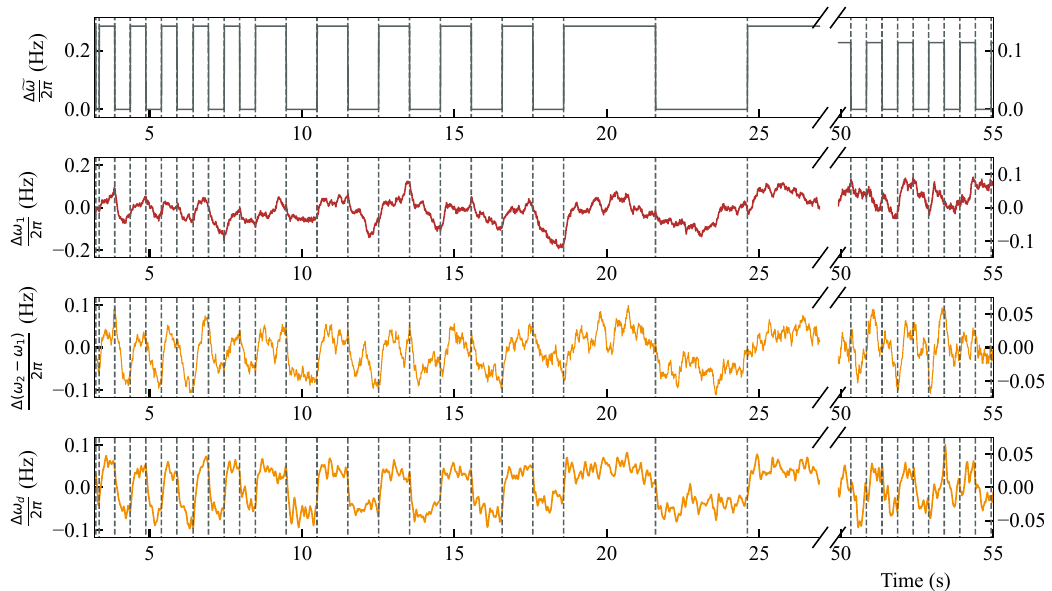}
 \caption{Post-processed cancellation vs non-linear transduced signal for one-on-one comparison}
 \label{fig:Sup_transducedvspost-processing}
\end{figure*}

Turning our attention to the signal at $\omega_\mathrm{d}$, we can easily see the correlation with $U$ due to the suppressed frequency fluctuations. Indeed, we obtain $\rho(U, \omega_\mathrm{d}) = 0.84$, which is \SI{65}{\percent} higher than $\rho(U, \omega_1)$. This measurement confirms that small frequency changes in $\omega_1$ are detected more sensitively by measuring $\omega_\mathrm{d}$ than $\omega_1$ itself. In direct comparison with the post-processing difference signal ( $\omega^\mathrm{cal}_\mathrm{d} = \omega_2-\omega_1$) to calculate $\rho(U, \omega^\mathrm{cal}_\mathrm{d}) =  0.43$. \figureref{fig:Sup_transducedvspost-processing} highlights the visual comparison to the post-processed data.

\begin{figure*}[!tbp]
 \includegraphics[]{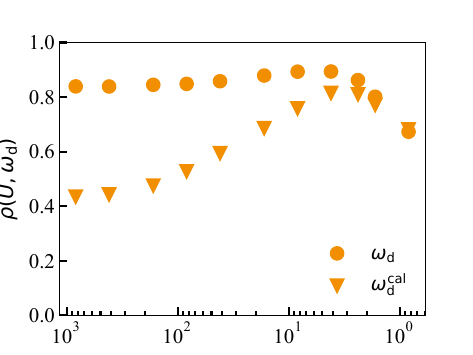}
 \caption{Normalised covariance between the transduced and post-processed $\omega_\text{d}$ as a function of averaging which translates into an effective sampling rate. The left-most datapoints corresponds to \SI{837}{Samples/\second}, i.e. without any averaging.}
 \label{fig:Sup_cov_averaging}
\end{figure*}

Now we return to calculating this figure of merit as a function of averaging or smoothing of the raw data, by convolving the data with a box filter of varying lengths (\url{https://stackoverflow.com/a/26337730}). As expected, for all sampling rates, the transduced data outperforms the post-processed data. This also comes from the fact that $\omega_2$ is not a PLL-tracked signal but a phase to frequency calculated signal~\cite{Besic2023ResonanceResonators}. We perform the same experiment where we instead track $\omega_1$ and $\omega_2$ and calculate $\omega_\mathrm{d}$ from phase. Even in this case, in spite of being limited by the differential calculation which lowers $\rho$ at higher sampling rates, the transduced signal shows a better performance than the post-processed signal. In both cases, as the effective sampling rate approaches \SI{10}{samples/\second} or lower, only the contributions from the $S_\omega$ branch in the f-PSD remain, hence their performance converges.

\section{Reproducibility on several devices}
Several measurements shown in the main text have been repeated over multiple devices to verify the reproducibility of this claim. Below, you find measurements on two such devices. The agreement among all three datasets confirms the reproducibility of the stabilization mechanism and demonstrates that the observed effects are robust across nominally similar devices.

\subsection{Device 53}
\begin{figure*}[!tbp]
 \includegraphics[]{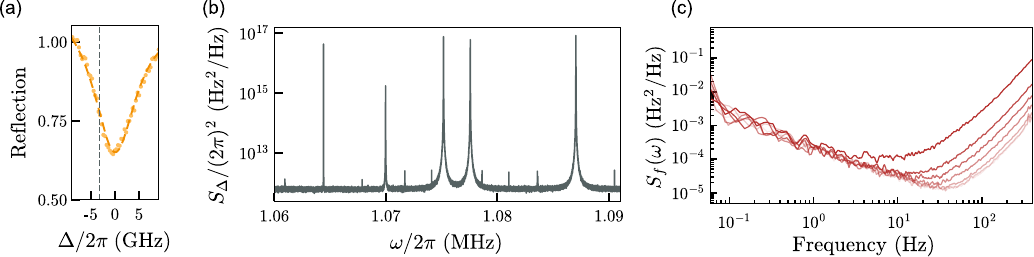}
 \caption{(a)~Normalized reflection trace of the optical mode used in the experiment. Yellow circles: measured reflection as a function of laser wavelength. Yellow dashed line: fit of the optical reflection. Vertical dashed line: cavity detuning used in experiments. (b)~Displacement PSD of the calibrated optical frequency noise. The peaks correspond to the thermomechanical noise stemming from the perimeter modes of the polygon resonator. Modes marked with red and blue are used in the experiment. (c)~Mechanical frequency noise PSD of the high-$Q$ mode at different optical powers.}
 \label{fig:SI_Fig_1_53}
\end{figure*}

\begin{figure*}[!tbp]
 \includegraphics[]{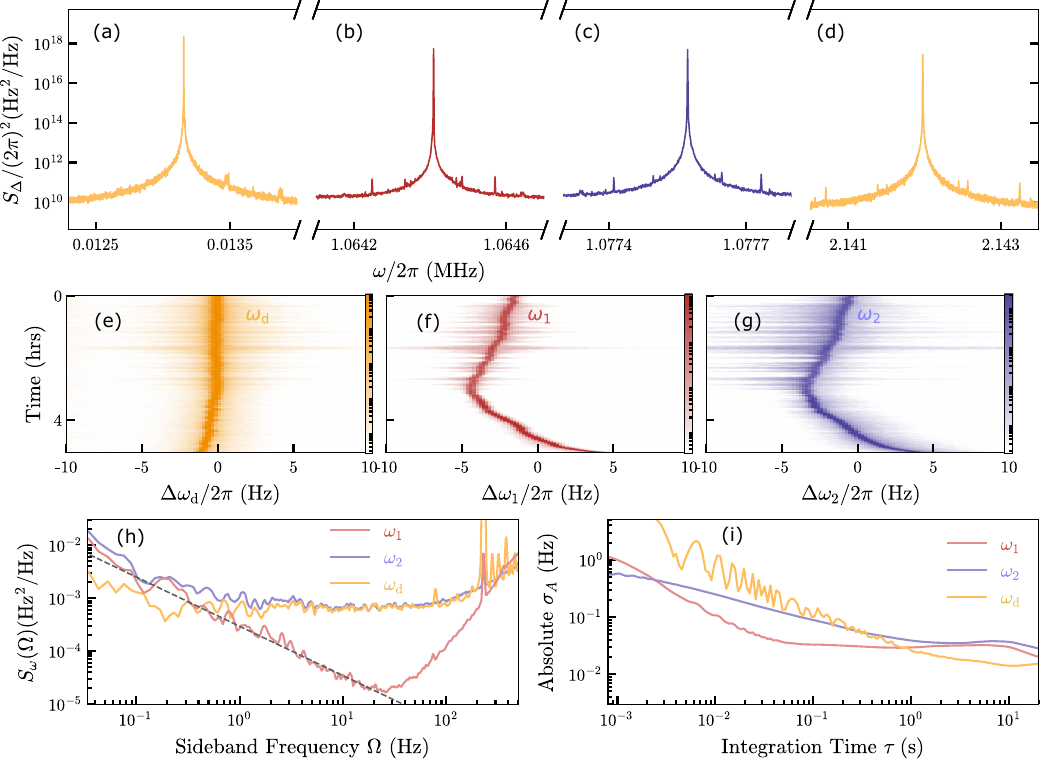}
 \caption{Properties of signal at $\omega_\mathrm{d} = \omega_2 - \omega_1$. Displacement PSD of demodulated signal around (a)~$\omega_\mathrm{d} = \omega_2 - \omega_1$ and (c)~$\omega_s = \omega_1 + \omega_2$. The mechanical modes at $\omega_1$ (red) and  $\omega_2$ (blue) are driven with separate PLLs. (b) Thermal mechanical spectrum without external drive, same as in \figref{fig:SI_Fig_1_53} (b) to highlight the modes $\omega_1$ and $\omega_2$. (d)-(f)~Spectograms of the difference signal and the displacements of the OOP and the IP modes. The color bar corresponds to the normalized PSD. (g)~Double-sided frequency-PSD for $\omega_1$, $\omega_2$, and $\omega_\mathrm{d}$ spanning \SI{150}{\second} with a data rate of \SI{14.3}{\kilo\hertz}. We use the PLLs for estimating $\omega_1$ and $\omega_2$, while  $\omega_\mathrm{d}$ is calculated using phase-to-frequency conversion~\cite{Besic2023ResonanceResonators}. (h)~Absolute Allan deviation $\sigma$ in \SI{}{\hertz} calculated using the same traces as in (h).}
 \label{fig:SI_Fig_2_53}
\end{figure*}

\begin{figure*}[!tbp]

 \includegraphics[]{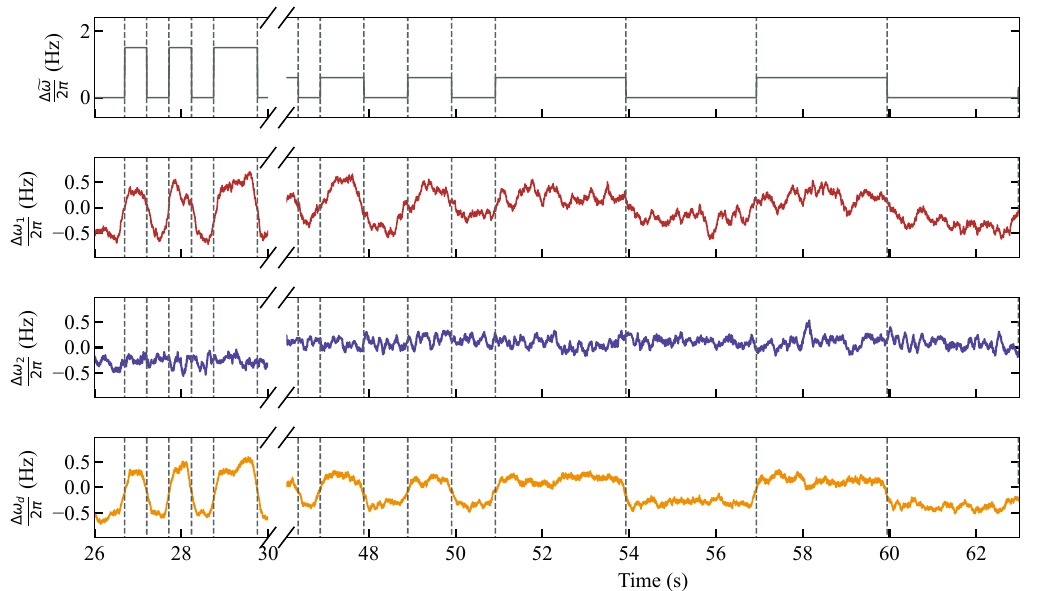}
 \caption{Measuring frequency jumps with high precision at $\omega_\mathrm{d}$. (a)~Schematic illustration of the effect of frequency jumps $\Delta\omega$ acting on $\omega_1$, but not on $\omega_2$. (b)~Expected frequency jumps $\Delta\widetilde{\omega}$ when a dispersive feedback on mode $\omega_1$ is toggled on and off. The two modes at $\omega_1$ and $\omega_2$ are driven by separate PLLs. Frequency shifts (c)~$\Delta\omega_1$, (d)~$\Delta\omega_2$, and (e)~$\Delta\omega_\mathrm{d}$ are simulataneously measured as a function of time.}
 \label{fig:SI_Fig_3_53}
\end{figure*}

We repeated the full set of measurements on a second optomechanical resonator of identical design. The cavity measurement and detuning~\figref{fig:SI_Fig_1_53}(a), thermal PSD~\figref{fig:SI_Fig_1_53}(b), effect of optical power on the frequency-PSD~\figref{fig:SI_Fig_1_53}(c) are shown. On driving the two modes $\omega_1$ and $\omega_2$, the voltage PSDs~\figref{fig:SI_Fig_2_53}(a-d), the thermal drifts~\figref{fig:SI_Fig_2_53}(e-g) and the frequency PSD~\figref{fig:SI_Fig_2_53}(h) and Allan deviation analysis~\figref{fig:SI_Fig_2_53}(i) are showcased. We also perform the experimental PLL-based frequency-tracking with feedback frequency jumps~\figref{fig:SI_Fig_3_53} to display the same qualitative response as in the primary device, including the characteristic frequency shift. 

\makeatletter
\par
\vspace{-1em}
\twocolumngrid
\makeatother


\subsection{Device 26 on a different chip}
\begin{figure*}[!tbp]
 \includegraphics[]{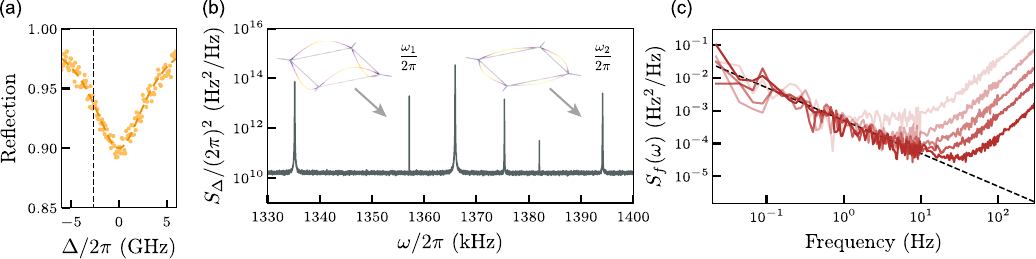}
 \caption{(a)~Normalized reflection trace of the optical mode used in the experiment. Yellow circles: measured reflection as a function of laser wavelength. Yellow dashed line: fit of the optical reflection. Vertical dashed line: cavity detuning used in experiments. (b)~Displacement PSD of the calibrated optical frequency noise. The peaks correspond to the thermomechanical noise stemming from the perimeter modes of the polygon resonator. Modes marked with red and blue are used in the experiment. (c)~Mechanical frequency noise PSD of the high-$Q$ mode at different optical powers.}
 \label{fig:SI_Fig_1_26}
\end{figure*}

\begin{figure*}[!tbp]
 \includegraphics[]{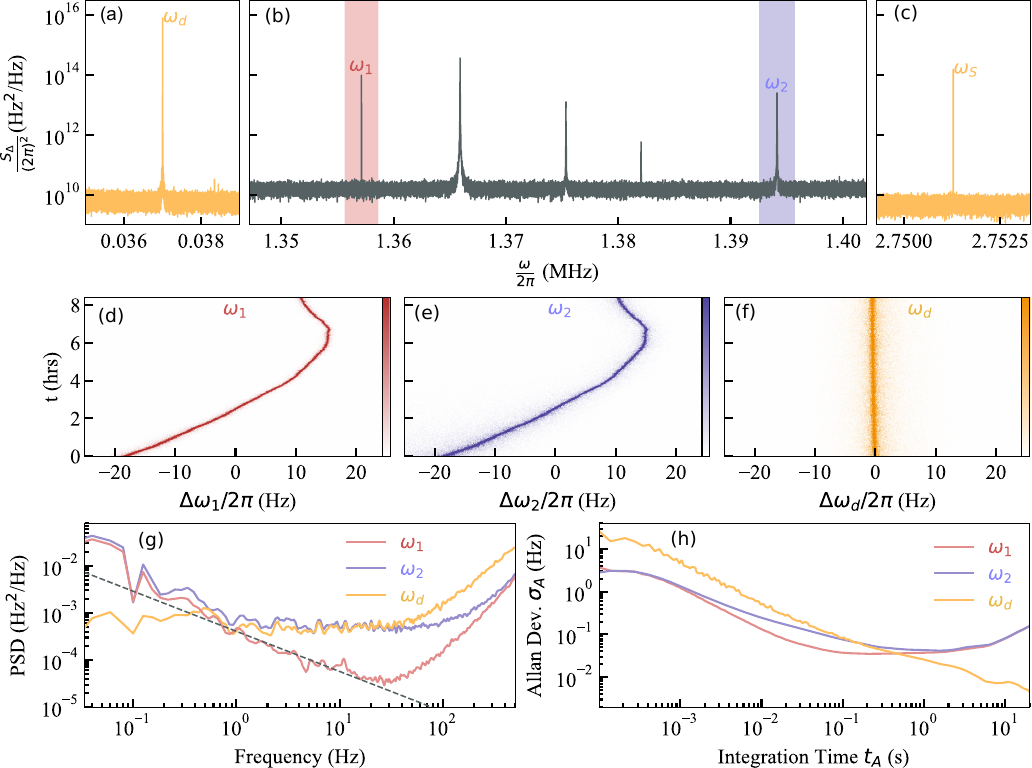}
 \caption{Properties of signal at $\omega_\mathrm{d} = \omega_2 - \omega_1$. Displacement PSD of demodulated signal around (a)~$\omega_\mathrm{d} = \omega_2 - \omega_1$ and (c)~$\omega_s = \omega_1 + \omega_2$. The mechanical modes at $\omega_1$ (red) and  $\omega_2$ (blue) are driven with separate PLLs. (b) Thermal mechanical spectrum without external drive, same as in \figref{fig:SI_Fig_1_26} (b) to highlight the modes $\omega_1$ and $\omega_2$. (d)-(f)~Spectograms of the difference signal and the displacements of the OOP and the IP modes. The color bar corresponds to the normalized PSD. (g)~Double-sided frequency-PSD for $\omega_1$, $\omega_2$, and $\omega_\mathrm{d}$ spanning \SI{150}{\second} with a data rate of \SI{14.3}{\kilo\hertz}. We use the PLLs for estimating $\omega_1$ and $\omega_2$, while  $\omega_\mathrm{d}$ is calculated using phase-to-frequency conversion~\cite{Besic2023ResonanceResonators}. (h)~Absolute Allan deviation $\sigma$ in \SI{}{\hertz} calculated using the same traces as in (h).}
 \label{fig:SI_Fig_2_26}
\end{figure*}

We characterise yet another device on a different chip with an identical design but different dimensions. The characterisation of the device is provided in~\figref{fig:SI_Fig_1_26}. The improved performance at $\omega_\text{d}$ is showcased in \figref{fig:SI_Fig_2_26}. 

\makeatletter
\par
\vspace{-1em}
\twocolumngrid
\makeatother

\end{appendix}
\end{document}